\documentstyle[12pt,aasms4]{article}
\journalid{337}{10 Mars 1997}

\def\mic{\ifmmode {\,\mu{\rm m}} \else $\,\mu{\rm m}$\fi}       
\def\kms{\ifmmode {\,{\rm km\,s^{-1}}}                          
	\else {\hbox{$\,$ {\rm km$\,$s$^{\rm -1}$}}}\fi}
\def\mo {\ifmmode {\,{\it M}\odot} \else $\,M$\odot\fi} 
\def\lo {\ifmmode {\,{\it L}\odot} \else $\,L$\odot\fi} 
\def\my {\ifmmode {\,{\it M}\solar\,{\rm yr^{-1}}}              
	\else {$\,M$\solar$\,$yr$^{\rm -1}$}\fi}
\def\cmm#1{\ifmmode {\,{\rm cm^{-#1}}}          
	\else \hbox{$\,${\rm cm$^{\rm -#1}$}}\fi}
\chardef\isp="10\def\i{\'\isp}		
\def\as {\ifmmode {^{\scriptscriptstyle\prime\prime}}           
	\else $^{\scriptscriptstyle\prime\prime}$\fi}
\def\am {\ifmmode {^{\scriptscriptstyle\prime}}                 
	\else $^{\scriptscriptstyle\prime}$\fi}
\def\deg {\ifmmode^\circ\else$^\circ$\fi}                       
\def\raw {\ifmmode\rightarrow\else$\rightarrow$\fi}             
\def\x {\ifmmode\times\else$\times$\fi}                         
\def\gsim {\ifmmode {\buildrel>\over\sim}               
	\else {\lower.6ex\hbox{$\buildrel>\over\sim$}}\fi}
\def\lsim {\ifmmode {\buildrel<\over\sim}               
	\else {\lower.6ex\hbox{$\buildrel<\over\sim$}}\fi}
\def\ra[#1 #2 #3.#4]{ #1$^{\rm h}$#2$^{\rm m}$#3$^{\rm s}$.#4}  
\def\dec[#1 #2 #3.#4]{ #1\deg#2\am#3{\as}.#4}             
\def\rax[#1 #2 #3]{RA: #1$^{\rm h}$#2$^{\rm m}$#3$^{\rm s}$}
\def\decx[#1 #2 #3]{Dec:#1\deg#2\am#3\as}          
\def\h2{\rm H$_2$}                      
\lefthead{Fuente A., Mart{\i}n-Pintado J.}
\righthead{ISO observations toward NGC 7023...}

\begin{document}

\title{ISO observations toward the reflection nebula NGC 7023: A 
nonequilibrium ortho- to para-H$_2$ ratio\footnotemark[1] 
}

\author{A. Fuente, J. Mart{\i}n-Pintado, N.J. Rodr{\i}guez-Fern\'andez, 
A. Rodr{\i}guez-Franco, P. de Vicente}
\affil{Observatorio Astron\'omico Nacional (IGN), Campus Universitario,
Apdo. 1143, E-28800 Alcal\'a de Henares (Madrid), Spain }

\author{D. Kunze}
\affil{Max-Planck Institut f\"ur Extraterrestrische Physik, Postfach
1603, D-85748 Garching, Germany }

\authoremail{fuente@oan.es}

\footnotetext[1]{Based on observations with ISO, an ESA project 
with instruments funded by ESA Member States (especially the PI
countries France, Germany, the Netherlands and the United Kingdom)
and with participation of ISAS and NASA}

\begin{abstract}
We have observed the S(0), S(1), S(2), S(3), S(4) and S(5) rotational lines 
of molecular hydrogen (H$_2$) towards the peak of the photodissociation
region (PDR) associated with the
reflection nebula NGC 7023. The observed H$_2$ line ratios show that they arise
in warm gas with kinetic temperatures $\sim$ 300 - 700 K. However, the data
cannot be fitted by an ortho- to para- (OTP) ratio of 3.
An OTP ratio in the range $\sim$ 1.5 - 2 is necessary to explain our
observations. This is the first detection of a non-equilibrium OTP
ratio measured from the H$_2$
pure-rotational lines in a PDR. The existence of a dynamical PDR is discussed
as the most likely explanation for this low OTP ratio.
\end{abstract}

\keywords{ISM: abundances --- ISM: individual (NGC 7023) --- reflection
nebulae --- stars: individual (HD 200775) --- stars: pre-main-sequence
--- infrared: ISM: lines and bands}

\section{Introduction}
The Short Wavelength Spectrometer (SWS) on board the Infrared Space
Observatory (ISO) has allowed us, for the first time, to observe the
pure rotational spectrum of H$_2$ and study the OTP 
ratio in warm (T$\sim$ 100 - 1000 K) regions.
Previous to ISO observations, the OTP ratio of  H$_2$ had been 
estimated 
towards regions heated by UV radiation and shocks  
to temperatures $>$ 2000 K 
using the relative strengths of the 
vibrational lines. In outflow regions where shock excitation
is the main heating mechanism, observations of the vibrational 
lines typically reveal OTP ratios of $\sim$ 3 (Smith,
Davis \& Lioure 1997). This is the expected value in thermodynamic 
equilibrium for gas with temperatures $>$ 200 K. 
However, towards the regions heated mainly by the UV radiation 
(PDRs), the vibrational lines give OTP values in the range 1.5 - 2
(see e.g. Chrysostomou et al. 1993). 
The pure rotational lines of H$_2$ have been detected by ISO toward
several Galactic regions (S140: Timmermann et al. 1996; Cepheus A West:
Wright et al. 1996; BD+40$\deg$4124: Wesselius et al. 1996; 
HH54: Neufeld et al. 1998). 
The observations towards all these regions except HH54
are consistent with the
H$_2$ rotational lines arising in gas with temperature $>$ 200 K
and with an OTP ratio of 3. This value is not consistent with those
derived from the vibrational lines in dense PDRs 
but is in agreement
with theoretical predictions.
Recently Sternberg \& Neufeld (1999) have argued that the low
OTP ratios measured from vibrational lines 
do not represent the actual OTP ratio in PDRs
but it is simply a consequence of the optical depth effects in the fluorescent
pumping of the vibrational lines.
The non-equilibrium OTP ratio measured towards HH54
has been interpreted as arising in hot-shocked excited gas that has not
reached the equilibrium yet. In this interpretation, the observed 
OTP ratio would be the legacy of an earlier stage in the thermal history of the
gas when the gas temperature was 90~K. 

In this {\it Letter} we report the observations of the H$_2$ rotational lines
towards the prototypical PDR associated with the reflection nebula
NGC 7023. An OTP ratio in the range  1.5 - 2 is derived from  
our observations. This is the first 
detection of an OTP ratio  $<$ 3 in a PDR based on the
pure H$_2$ rotational lines.
\section{Observations}
The observations of the S(0), S(1), S(2), S(3), S(4) and S(5) rotational
lines of H$_2$ towards the peak of the PDR associated with 
NGC 7023 (R.A. (2000):\ra[21 01 32.5], Dec(2000):\dec[68 10 27.5]). 
were made using the Short-Wavelength Spectrometer (SWS)
(de Grauuw et al. 1996) on board the Infrared Space Observatory
(ISO) (Kessler et al. 1996) during revolution 514 with a total on-target time
of 5149 s. At the spectral resolution of this mode 
($\lambda$/$\Delta$$\lambda$ $\sim$ 1000-2000) all the observed lines are
unresolved. Data reduction are carried out with version 7.0 of the Off Line
Processing routines and the SWS Interactive Analysis at the ISO Spectrometer 
Data Center at MPE. Further analysis has been made using the ISAP software
package. The uncertainities in the calibration are of 15 \%,
25 \%, 25 \%, 25 \%, 20\% and 30\% for the S(5), S(4), S(3), S(2), S(1) and
S(0) lines respectively (Salama et al. 1997). 
 
\section{H$_2$ rotational lines}
In Fig. 1 we present the spectra of
the S(0), S(1), S(2), S(3), S(4) and S(5) H$_2$ lines towards the 
peak of the PDR associated with NGC 7023. The observed intensitites 
and some interesting observational parameters are presented in Table 1. 
The data have been corrected for dust attenuation 
using the extinction curve of Draine \& Lee (1984), 
and a value of 0.43
for the dust opacity at 0.55 $\mic$.
This value has 
been derived from the ISO LWS01 spectrum (Fuente et al.
1999, hereafter paper II).
The extinction for the
S(0), S(1), S(2), S(3), S(4) and S(5) lines is very small and
amounts to 0.01, 0.02, 0.02,
0.04, 0.02 and 0.01 respectively.
Fig. 2 shows the rotational diagram for the H$_2$ corrected for extinction
effects and assuming extended emission. (Note that the errors in Fig. 2
are entirely dominated by the calibration uncertainities.)

This diagram shows that the ortho-H$_2$ levels 
have systematically lower N$_u$/g$_u$ values
(where N$_u$ and g$_u$ are the column densities and 
degeneracies of the upper levels of the transitions) 
than the adjacent J--1 and J+1 para-H$_2$ levels
producing a $``$zig-zag" distribution. 
In fact, the ortho-H$_2$ levels seem to define
a curve which is offset from that of the para-H$_2$ levels (see Fig. 2).
This is the expected trend if the OTP ratio is
lower than 3. The offset between the two set of
data is systematic and seems to show the trend of 
being larger for the low energy levels (S(0) and S(1) transitions),
than for the high energy levels (S(2), S(3), S(4) and S(5) transitions).

The offset between the ortho- and para- H$_2$ curves is larger 
than the observational errors and 
cannot be due to the different apertures for the different lines.  
For the case of a point-like source we would have to correct by a factor
of 1.93 the intensity of the S(0) line and by a 
factor of 1.35 the intensities of the S(1) and S(2) lines. 
In this case, the offset between the para- and ortho- curves 
would increase.
Based on the spatial distribution of the HI(21 cm) line (Fuente et al. 1998) 
and those of the CII (158 $\mic$) and OI (63 $\mic$) lines 
(Chokshi et al. 1988,
paper II), we will assume in the following a beam filling factor of 1 
for all the observed H$_2$ lines. 
Although the exact shape of the rotational diagram depends
on the excitation conditions of the region,
the $``$zig-zag"  features cannot be
explained by any model which assumes an equilibrium OTP ratio. 
In thermodynamic equilibrium,
the rotational temperature between an ortho- (para-) H$_2$ level 
and the next J+1 level should always increase with
the energy of the upper level giving rise to a smooth curve in the
rotational diagram regardless of the temperature profile of
the region. A $``$zig-zag" distribution implies that for 
some pair of levels 
the rotational temperature decreases with the energy of the upper level.
One can only get this effect with a non-equilibrium OTP ratio.
 
We have compared our data with the models by Burton et al. (1992) 
in order to estimate the incident UV field, density and OTP ratio.
Burton et al. (1992) assumed a constant value of 3 for the OTP 
ratio in their calculations. The best 
fit to our data is for an incident UV field of G$_o$ = 10$^4$
in units of the Habing field and a density of n = 10$^6$ cm$^{-3}$, 
but it systematically underestimates
the intensity of the para-H$_2$ transitions and overestimates the
intensities of the ortho-H$_2$ transitions (see open triangles in Fig. 3). 
This is the expected behavior if one assumes
an OTP ratio of 3 and the actual OTP ratio in the region is $<$ 3. 
We have corrected the line intensities predicted by the model for
different values of the OTP ratio assuming that the total amount of H$_2$
molecules at a given temperature and the line ratios between levels 
of the same symmetry are not affected by the OTP ratio. These 
assumptions are correct for the low-J transitions where
the collisional excitation dominates.
In Fig. 3 we compare our data with the predicted diagram 
for an OTP ratio of 3  which correspond to the model
without any correction (open triangles), 
2 (open circles) and 1 (open squares).  
The diagram for an OTP ratio of 3 clearly does not fit any of 
our observational points. We find the best fit to our data with 
an OTP ratio of 2.
In this case, the predicted intensities for the S(1), S(2) and S(3) lines
are in agreement with the observed values.  
Only the predicted intensity for the S(0) line is not consistent
with the observations. To fit the intensity of the S(0) line it is
necessary to assume an OTP ratio of $\leq$ 1.5 (see the case OTP = 1 in
Fig. 3), but in this case, we will have
a worse fit for the S(1), S(2) and S(3) lines than for an OTP ratio of 2.   
As discussed in Section 4, this suggests a possible variation of 
the OTP ratio with the temperature.
The cooler gas emitting only in the S(0) line 
could have an OTP ratio of $<$ 1.5, while
the warmer gas emitting in the S(1), S(2) and S(3) lines
could have an OTP ratio of 2.
However, to establish this variation 
unambigously, it is necessary to have a 
very accurate knowledge of the excitation 
conditions in the region. To be conservative,
we will conclude that the OTP ratio  
is in the range 1.5 - 2 in this PDR. Martini et al. (1997) found an 
OTP ratio of 2.5$\pm$0.4 
using the near-IR H$_2$ vibrational lines towards their 
position 1 which is 20$''$ offset from ours. 
Taking into account that because of the optical depth effects in the
vibrational lines this value is just a lower limit to the
actual OTP ratio, it proves that the OTP ratio is close to 3 
for the gas with kinetic temperatures T$_k$ $>$ 2000 K.
This difference between the OTP ratios derived from the 
rotational and vibrational lines 
argues in favor of a variaton of
the OTP ratio with the kinetic temperature.  

Since the OTP ratio in this source is different from the equilibrium
value, the rotation temperature between an ortho- and a para- level
does not represent an estimate of the gas kinetic temperature. To
estimate the gas kinetic temperature we have calculated the rotation
temperature between levels of the same symmetry. 
For the para-H$_2$ levels, we have derived a rotation temperature
of $\sim$ 290 K from the S(0) and S(2) lines and of
$\sim$ 500 K from the S(2) and S(4) lines. For the ortho-H$_2$ levels, 
the derived temperature is $\sim$ 440  from S(1) and S(3) lines and
$\sim$ 700 K from the S(3) and S(5) lines. 
Based on these calculations we 
conclude that the OTP ratio is $\sim$ 1.5 - 2  in the
gas with  kinetic temperatures $\sim$ 300 - 700 K.
We derive a total H$_2$ column density of 5 10$^{20}$~cm$^{-2}$
assuming a rotation temperature of 290 K between the J = 0 and J = 2
para-levels and of 440 K between the J = 1 and J = 3 ortho-levels. 
 
\section{Discussion}
Based on the H$_2$ rotational lines data 
we have derived an OTP ratio in the range 
of 1.5 - 2 for the gas with kinetic temperatures $\sim$ 300 - 700 K
towards the reflection nebula NGC 7023. This is the second object with
a non-equilibrium OTP ratio measured from the H$_2$ pure rotational lines. 
The first non-equilibrium OTP ratio was
detected towards the outflow source HH54, and interpreted as arising in 
shock-heated gas that has not reached
the equilibrium yet. This interpretation is not plausible for NGC 7023.
The high dust temperature,T$_d$$\approx$ 40 K, and the detection of 
the SiII (34.8 $\mic$) and CII (157.7 $\mic$) lines (paper II) 
proves the existence of a PDR in this region. 
Although a shock component might also exist 
(see Martini et al. 1997), the heating is dominated by UV photons. 
It is then expected that most of the warm H$_2$ arises in the PDR.

A non-equilibrium value of the OTP ratio is not expected for the
physical conditions prevailing
in a dense PDR. The initial OTP ratio after 
the H$_2$ formation is very uncertain. Because of the large exothermicity 
of the formation process, it is expected to be 3 .
However, this OTP ratio can change if the H$_2$ molecules remain on
the grain surface long enough to reach the equilibrium at the dust
temperature. In this case, the OTP ratio after H$_2$ formation
will be that of the equilibrium at the grain temperature.  
After the ejection of the H$_2$ molecules to the gas phase, 
exchange reactions with H and H$^+$ change the OTP ratio 
until achieving the equilibrium at the gas temperature.
For the gas temperatures traced by the H$_2$ rotational
lines ($>$ 200 K), the equilibrium OTP ratio
is 3. One possibility to explain
the low OTP ratio observed in NGC 7023 is to suppose that
the OTP ratio in the H$_2$ formation is lower than 3,
and once in the gas phase, 
the H$_2$ molecules are destroyed before attaining the  
equilibrium value at the gas temperature. 
The dust temperature in a PDR with G$_o$ = 10$^4$
and n = 10$^6$ cm$^{-3}$ is $<$ 100 K. In particular, the dust temperature
measured towards the PDR peak in NGC 7023 using our LWS01 spectrum is
40 K which corresponds to an equilibrium OTP ratio of $\sim$ 0.1. 
The OTP conversion in the atomic region is dominated by H - H$_2$ collisions 
with a rate
of 10$^{-13}n$ s$^{-1}$ where $n$ is the atomic hydrogen density
(see Sternberg \& Neufeld 1999 and references therein).
The hydrogen density derived from the HI image published 
by Fuente et al. (1998) is $\sim$ 5 10$^3$ cm$^{-3}$. A similar value
($\sim$ 0.5 - 1 10$^4$ cm$^{-3}$)
was derived by Chokshi et al. (1988) based on the OI and CII lines. 
However, densities larger than 10$^5$ cm$^{-3}$ are derived from molecular
data (Fuente et al. 1996, Lemaire et al. 1996, Gerin et al. 1998).
The density of atomic hydrogen in the region  with T$_k$ $\sim$ 300 K
is expected to be about an order of magnitude lower than
the H$_2$ density. Then we assume an atomic hydrogen density of            
$\sim$ 10$^4$ cm$^{-3}$ in our calculations. 
With this value, the
OTP conversion rate due to H$_2$ - H collisions 
is 10$^{-9}$ s$^{-1}$.
At the cloud surface the unshielded photodissociation rate is 
$\sim$ 5 10$^{-11} G_\circ$ s$^{-1}$ $\sim$ 5 10$^{-7}$ s$^{-1}$, i.e.,
2 orders of magnitude larger than the OTP conversion rate.
Then, in the cloud surface, before self-shielding is important
(A$_v$ $\leq$ 0.3 mag, T$_k$$>$ 500 K), an OTP ratio lower than 3 
can be explained by assuming that the OTP ratio in the H$_2$ formation
is the equilibrium value at the grain temperature.     
Deeper into the molecular cloud, when H$_2$ is self-shielded,
one can consider that the  
H$_2$ destruction rate is similar to the H$_2$ reformation rate
which is given by $\sim$ 3 10$^{-17} n$ s$^{-1}$. In this region,
a lower limit to the OTP conversion rate is given
by the conversion rate due to H$_2$ -  H$^+$ collisions
which is  $\sim$ 10$^{-17} n$ s$^{-1}$ (Sternberg \& Neufeld 1999). 
The OTP conversion
rate is of the same order than the destruction rate and the 
OTP ratio is expected to be close to the equilibrium. 
According to these estimates one expects to have an OTP ratio close to 3
at low temperatures and a non-equilibrium OTP ratio at higher temperatures.
The contrary trend is derived from our observations.
The high energy S(1), S(2) and S(3) lines are fitted with an 
an OTP ratio of 2  while the low energy S(0) line is better fitted
with an OTP ratio of 1.5.

Another possibility to explain the non-equilibrium OTP value
in NGC 7023 is to consider the case of a dynamic PDR, i.e. the
dissociation front is advancing into the molecular cloud. In this case
we do not have to assume a non-equilibrium OTP ratio in
the H$_2$ formation. 
The PDR is being fed continuously by the cool 
gas of the molecular cloud 
in which the equilibrium value of the OTP ratio is lower than 3. 
This gas is heated by
the stellar UV radiation to temperatures $>$ 200 K but leaves the PDR  
before attaining the equilibrium OTP ratio at this temperature. 
In this scenario, 
the gas which is expected to have an OTP
ratio smaller than the equilibrium value is the gas that has 
more recently been 
incorporated into the PDR, which is also the gas at
lowest temperature.
The OTP ratio is expected to increase and reach values
close to 3 at high temperatures.
This behavior is consistent with the
trend observed in our data. 
As discussed above, 
the OTP H$_2$ conversion
is mainly due to H - H$_2$ collisions with a conversion rate
of $\sim$ 10$^{-9}$ s$^{-1}$.
To have a significant fraction of the gas with a non-equilibrium
OTP ratio the photodissociation front must advance about
1 - 2 mag in the conversion time, i.e.,
it must penetrate into the molecular cloud
at a velocity of $\sim$ 10$^7 n^{-1}$ kms$^{-1}$. 
Skinner et al. (1993) proposed the existence
of an anisotropic ionized wind associated with this star based on radio
continuum observations. Later, Fuente et al. (1998) detected 
an HI outflow with a velocity of
$\sim$ 7.5 kms$^{-1}$. They proposed that this outflow is formed
when the gas that has been photodissociated by the UV radiation,
is accelerated by the stellar winds along the walls of a biconical
cavity. The velocity of the 
HI outflow cannot account for the velocity
at which the photodissociation front must advance into the
molecular cloud to have an OTP ratio lower than 3
unless the OTP conversion rates are severely
overestimated (by a factor of $>$ 10) and/or the density of the
outflow is $>$ 10$^5$ cm$^{-3}$. In spite of 
this problem, a photodissociation front penetrating into the
cloud because of the outflow seems to be the
most plausible explanation for the
low OTP ratio measured in this region. 
The existence of an outflow can also explain 
the difference between this region and other PDRs like S140 in which 
an OTP ratio of 3 has been
derived from the H$_2$ rotational lines. 

\acknowledgments

We would like to thank the referee for his/her helpful comments.
This work has been partially supported by the Spanish 
DGES under grant number PB93-0048 and the Spanish PNIE under
grant number ESP97-1490-E. 
N.J.R-F  acknowledges Conserjer{\i}a de Educaci\'on y Cultura de la
Comunidad de Madrid for a pre-doctoral fellowship.

\newpage

\figcaption[letfig1.ps]{ Spectra of the H$_2$ S(0),S(1),S(2),S(3),S(4) 
and S(5) towards the peak of the PDR associated with NGC 7023
(R.A. (2000):\ra[21 01 32.5], Dec(2000):\dec[68 10 27.5]). The S(1), S(2),
S(3), S(4), and S(5) lines are offset by 4,8,12,18 and 22 W cm$^{-2}$ 
$\mic^{-1}$ respectively.   
\label{fig1}}

\figcaption[letfig2.ps]{ Rotational diagram of the H$_2$ lines towards
peak of the PDR associated with NGC 7023. The dashed-line connect the
para- H$_2$ levels while the continuous line connect the ortho- H$_2$
levels. Note that there is an offset between the two curves and that 
this offset decreases for high temperatures.
\label{fig2}}

\figcaption[letfig3.ps]{ Comparison between the observational data
(filled squares) and the rotational diagram predicted by the model
by Burton et al. (1992) for G$_\circ$ = 10$^4$ and n = 10$^6$ cm$^{-3}$
corrected for different values of the OTP ratio, 3 (open triangles),
2 (open circles) and 1(open squares).
\label{fig3}}

\newpage
\begin{table*}
\caption{Observational parameters \label{tbl-1}}
\begin{tabular}{l ccc }
\\ \hline
\multicolumn{1}{l}{Line} &
\multicolumn{1}{l}{($\lambda$)} &
\multicolumn{1}{c}{beam} &
\multicolumn{1}{c}{Intensities}  \\
\multicolumn{1}{l}{} &
\multicolumn{1}{l}{($\mic$)} &
\multicolumn{1}{c}{($''$x$''$)} &
\multicolumn{1}{c}{(erg cm$^{-2}$ sr$^{-1}$)} \\
 \\ \hline
H$_2$ S(5) & 6.909 & 14 x 20 & 26.4$\pm$3.0 10$^{-5}$ \\
H$_2$ S(4) & 8.025 & 14 x 20 & 15.0$\pm$0.2 10$^{-5}$ \\
H$_2$ S(3) & 9.665 & 14 x 20 & 40.8$\pm$1.3 10$^{-5}$  \\
H$_2$ S(2) & 12.279 & 14 x 27 & 24.0$\pm$3.3 10$^{-5}$ \\  
H$_2$ S(1) & 17.035 & 14 x 27 & 21.4$\pm$0.7 10$^{-5}$  \\
H$_2$ S(0) & 28.219 & 20 x 27 & 3.4$\pm$1.0  10$^{-5}$  \\ \hline
\end{tabular}
\end{table*}
\clearpage


%
%
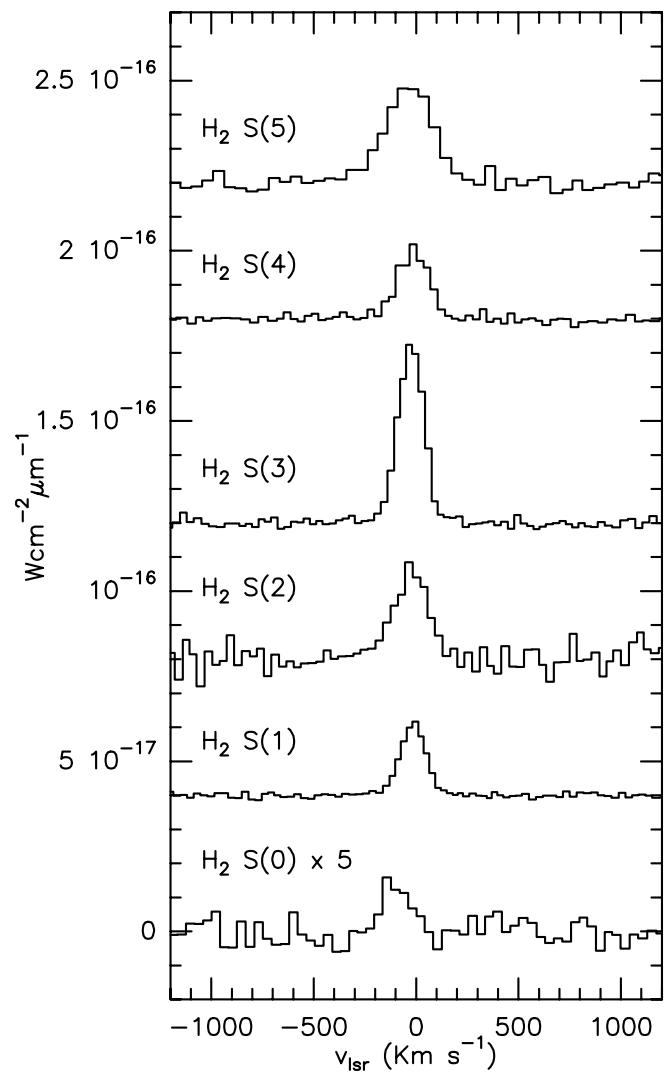
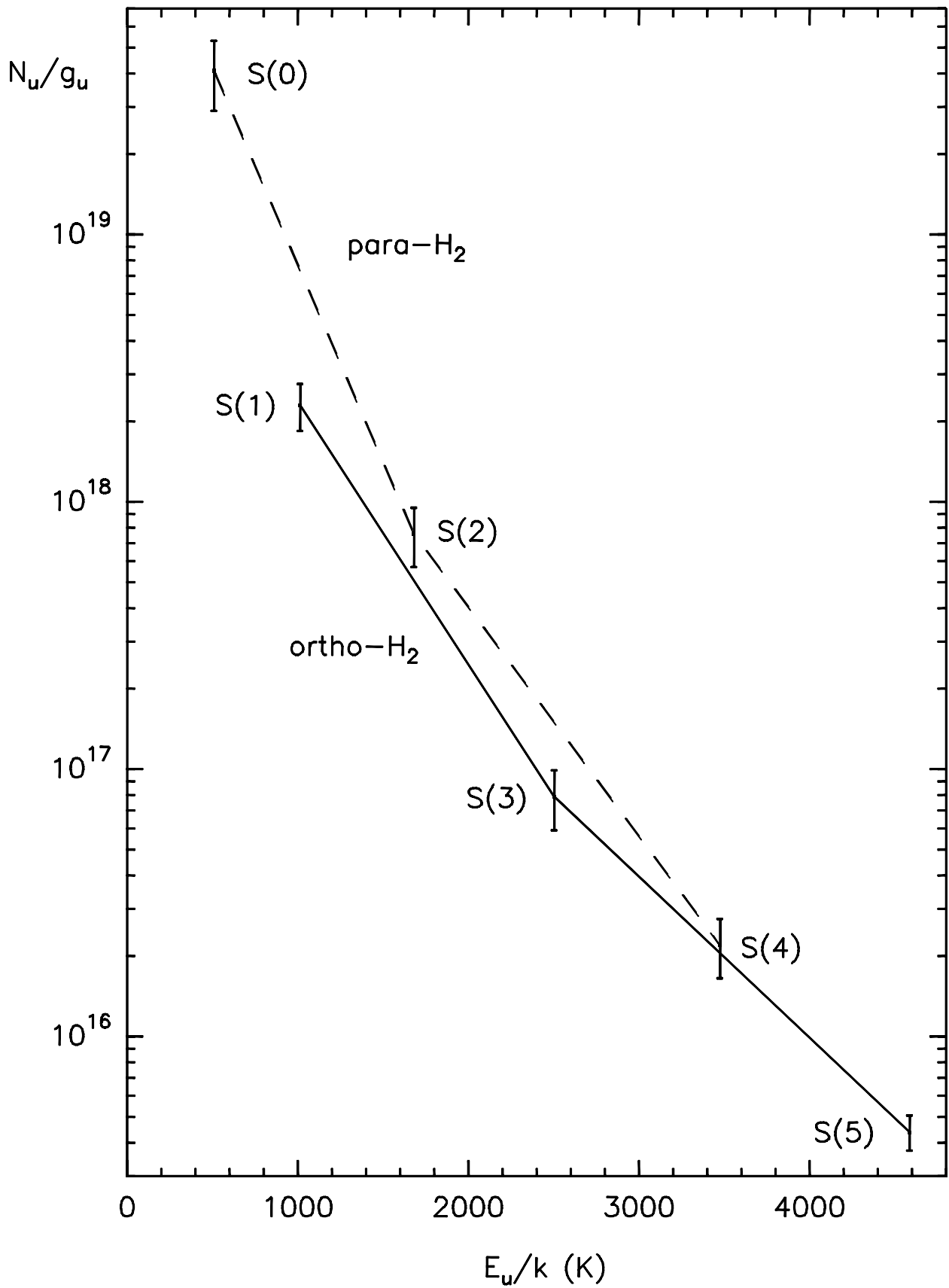
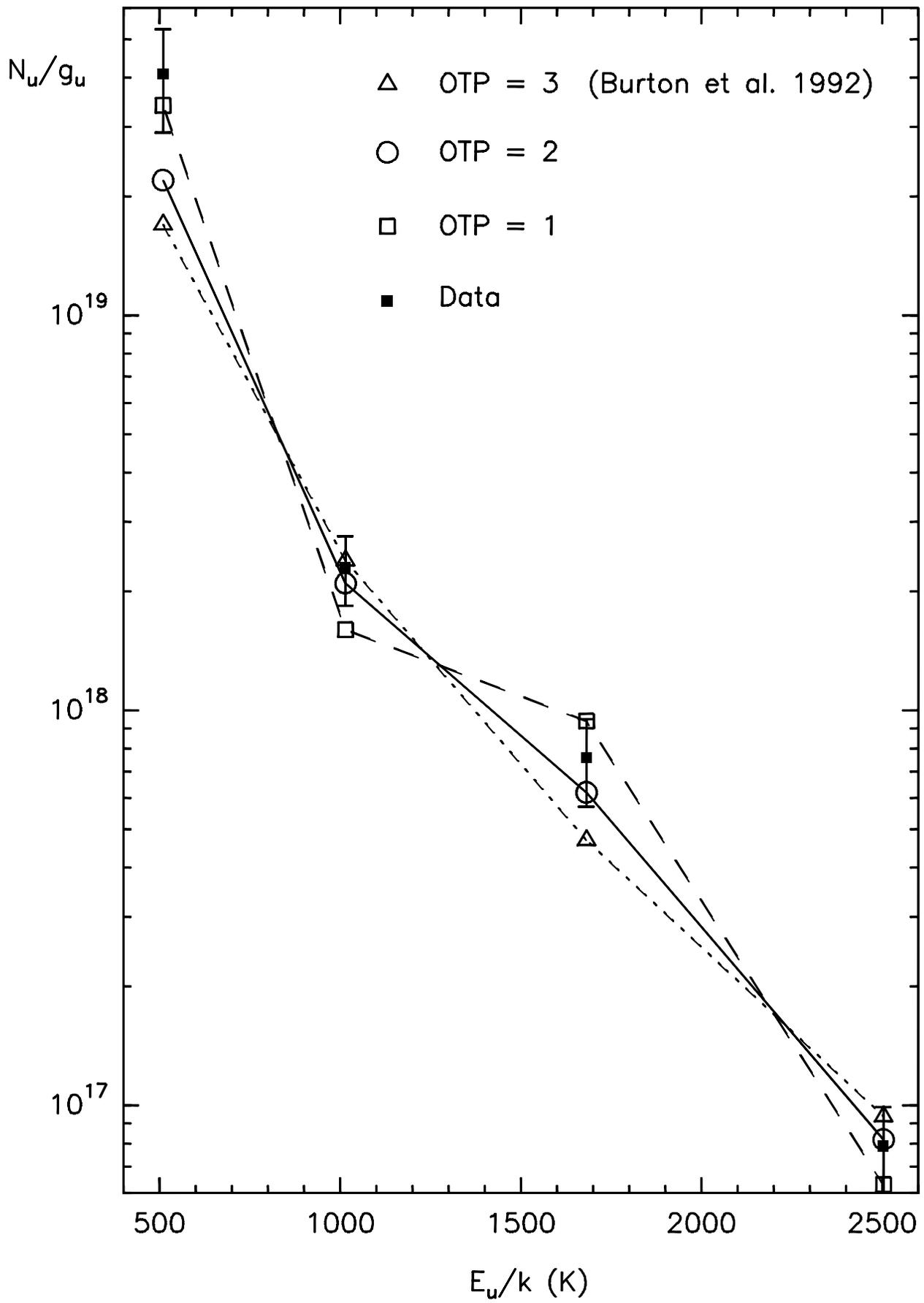
\end{document}